\documentclass[11pt]{article}
\usepackage{moriond,epsfig}

\def\Journal#1#2#3#4{{#1} {\bf #2}, #3 (#4)}

\def\apj{{ApJ}}%
\def\apjl{{ApJ}}%
\def\mnras{{MNRAS}}%
\def\aj{{AJ}}%
\def\araa{{ARA\&A}}%

\def\be{\begin{equation}}
\def\ee{\end{equation}}
\def\bea{\begin{eqnarray}}
\def\eea{\end{eqnarray}}

\def\Ha{\ifmmode \mathrm{H}{\alpha}\else H$\alpha$\fi}
\def\sun{\odot}
\def\Msol{\ifmmode \mathrm{M}_{\sun}\else M$_{\sun}$\fi}
\def\arcmin{\mbox{$^\prime$}}
\def\arcsec{\mbox{$^{\prime\prime}$}} 
\begin{document}
\vspace*{4cm}
\title{STAR FORMATION AT REDSHIFT ONE: PRELIMINARY RESULTS FROM AN \Ha\ SURVEY}

\author{Michelle Doherty$^1$,
Andrew Bunker$^{1,2}$,
Robert Sharp$^{1,3}$,
Gavin Dalton$^4$,
Ian Parry$^1$,
Ian Lewis$^4$,
Emily MacDonald$^4$, Christian Wolf$^4$}
\address{ $^1$Institute of Astronomy, Madingley Road, Cambridge, CB3~0HA, UK (\texttt{e-mail:md@ast.cam.ac.uk})\\
 $^2$School of Physics, University of Exeter, Stocker Road, Exeter, UK\\
 $^3$Anglo-Australian Observatory, Sydney, Australia\\
 $^4$Astrophysics, NAPL, Keble Road, Oxford, OX1~3RH, UK}

\maketitle\abstracts{
We present the first successful demonstration of
multi-object near-infrared spectroscopy on high redshift
galaxies. Our objective is to address the true star formation
history of the universe at $z\sim1$, a crucial epoch which some
have suggested is the peak of star formation activity.
By using H$\alpha$ -the same robust star formation indicator
used at low-z - redshifted into the J- and H-bands, we can
trace star formation without the systematic uncertainties
of different calibrators, or the extreme dust extinction
in the rest-UV, which have plagued previous efforts.
We are using the instrument CIRPASS (the Cambridge Infra-Red
PAnoramic Survey Spectrograph), in multi-object mode, which has been successfully
demonstrated on the Anglo-Australian Telescope (AAT) and
the William Herschel Telescope (WHT). CIRPASS has 150 fibres
deployable over $\sim40\arcmin$ on the AAT and $\sim15\arcmin$ on the
WHT. Here we present preliminary results from one of our
fields observed with the WHT: H$\alpha$ detections of z$\sim$1
galaxies in the Hubble Deep Field North.
}
\section{Introduction}
\subsection{Background}

An important issue in studies of galaxy formation and evolution is determining the peak epoch of star formation in the universe.  
There is clear evidence that the star formation rate (SFR) was higher in
the recent past - it is known to rise steeply to $z\sim1$ and perhaps to
peak around redshifts $1-2$. (e.g. Lilly et al. 1996, Tresse et
al. 2002). However there are large discrepancies between SFR estimates
obtained from different methods: different indicators have uncertain
relative calibration and are also differently affected by dust
extinction. There is a factor of $\sim10$ uncertainty in the star formation
history of the universe at $z\sim1$ (e.g. Hopkins et al. 2001), a
discrepancy which needs to be addressed by using a uniform tracer of star
formation across all redshift bins. The \Ha\ emission line is a reliable
tracer of the instantaneous SFR as the \Ha\ luminosity of a galaxy is
directly proportional to the ionising flux from its massive stars. It is
also relatively immune to metallicity effects and less susceptible to
extinction than the rest-UV. However, tracing \Ha\ to early epochs forces
a move to the NIR at $z>0.6$. Previous \Ha\ surveys in the NIR have so far
been restricted to small samples, as they have relied on longslit
spectroscopy which is inefficient for surveys, in terms of telescope
time (e.g. Glazebrook et al. 1999, Tresse et al. 2002, Erb et al. 2003). A large sample (several hundred) of galaxies is needed to address this matter properly.   

\subsection{An H$\alpha$ Survey with the CIRPASS multi-object spectrograph}

We are using the fibre-fed CIRPASS spectrograph (Parry et al. 2000), to
undertake an H$\alpha$ survey of galaxies at $z=0.7-1.5$ (Doherty et
al. 2004), in order to address the true star formation history of the
universe at this epoch. We aim to target several hundred galaxies. We have
carried out a successful pilot study for this project, performing multi-object spectroscopy of galaxies in the vicinity of the HDF-N. The instrument operates in the near-infrared, between 0.9 and 1.67$\mu$m and was used in multi-object mode, with 150 fibres of 1.1\arcsec\ diameter on the sky (comparable to the expected seeing convolved with typical galaxy profiles), deployable over 15\arcmin\ field of view. The plate was centroided to an accuracy better than 0.5\arcsec\ using six guidebundles centred on bright stars. A Hawaii 2K detector was used, and a grating of 831 lines/mm,
producing a dispersion of 0.95\AA/pixel. The full width at half maximum
(FWHM) of each fibre
extends over 2.7 pixels in both the spatial and spectral domain. The
wavelength coverage was 1726\AA, covering most of the J-band, and the
grating was tilted to set a central wavelength of $\lambda_c =
1.25 \mu$m.  The resolving power was $R =
\lambda/\Delta\lambda_{FWHM}\approx5000$. At this resolution the
background is very dark between the OH sky lines in the J- and
H-bands, allowing for high sensitivity to emission lines falling between
the skylines (which contaminate only about 10\% of the array). We used half
of the fibres on target objects and half on sky, to allow for good sky
subtraction. Thus with the capability of observing 75 objects
simultaneously, CIRPASS offers a huge multiplex advantage, ideal for
carrying out such a survey. 

\section{Initial Results: Bright \Ha\ galaxies in the HDF-N}

\begin{figure}
\begin{tabular}{c}
J1237063+6214027 \\
\psfig{figure=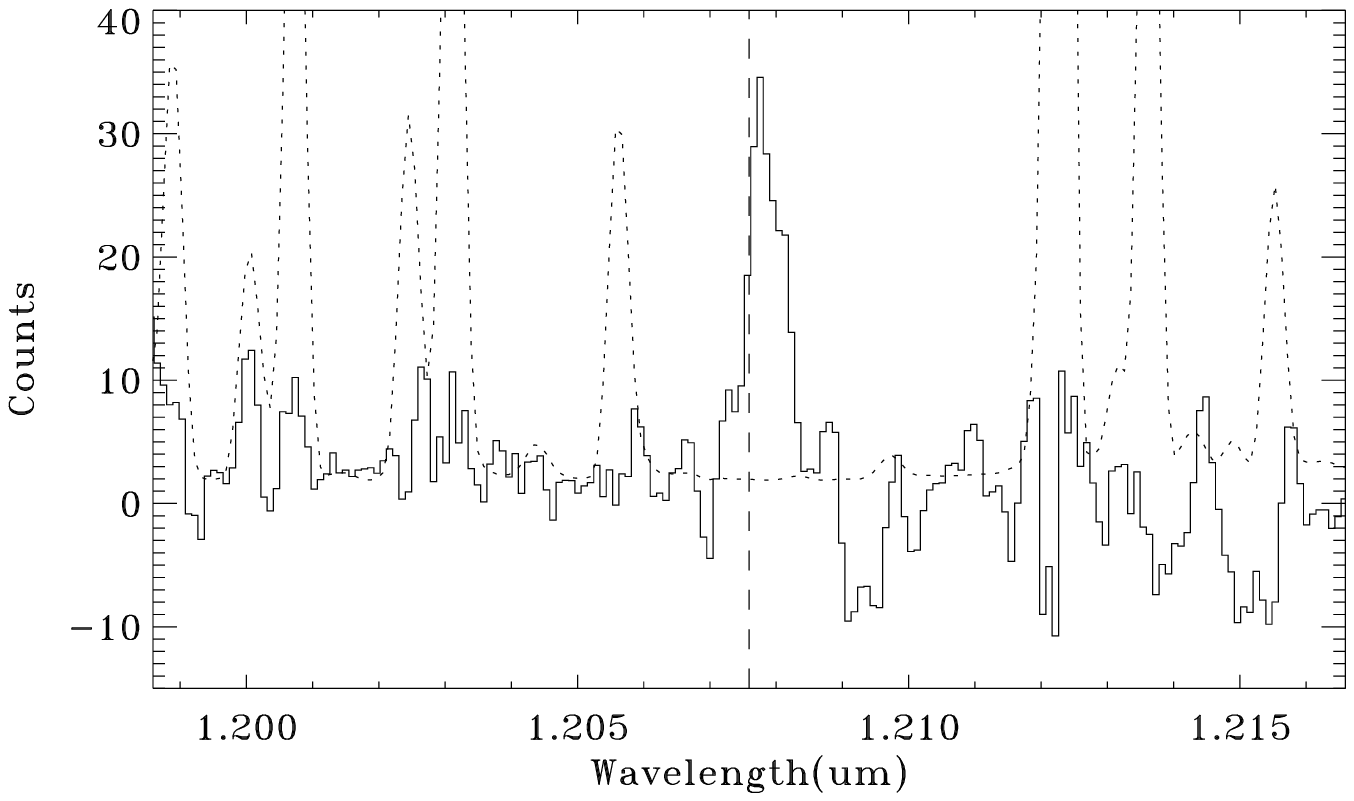,height=70mm} \\
J1237087+6211285 \\
\psfig{figure=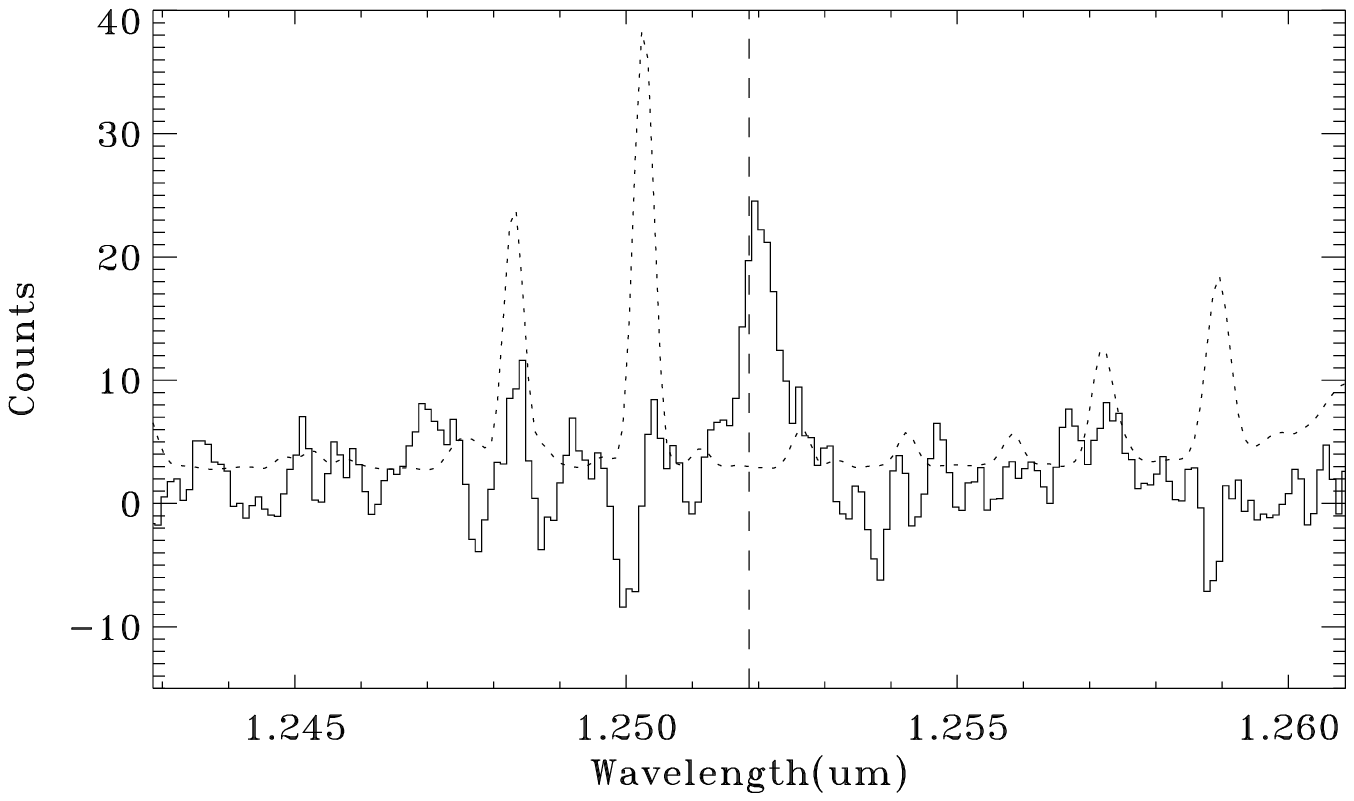,height=70mm}\\  
\end{tabular}
\caption{Example spectra for two objects where \Ha\ was detected at more
  than $5\sigma$. The night sky spectrum is overlaid in dotted lines for
  convenient identification of sky residuals which did not fully subtract
  out. The expected position of \Ha\ at the optical spectroscopic redshift
  (Cohen et al. 2000)
  is marked with a dashed line.  }
\label{spectra}
\end{figure}

We detected \Ha\ in a sub-sample of our targets in the HDF-N at $>5\sigma$
in just 3 hours of observing. Examples are shown in Figure~\ref{spectra}. In an average
fibre, between skylines, for a spectrally unresolved line (velocity FWHM $<60$ km s$^{-1}$)
we achieve a sensitivity of $7.2\times10^{-17}{\rm ~erg~s^{-1}~cm^{-2}}$ at
$5\sigma$ (in 3 hours, equivalent to a star forming rate of $5\,h^{-2}_{70}\,M_{\odot}\,{\rm yr}^{-1}$). The emission lines we detect are typically broader
than this, with velocity full width at half maximum of  $\sim100 -
250 {\rm ~km~s}^{-1}$, equivalent to $\sigma_{1D}=40-100 {\rm ~km~s}^{-1}$. Some
of this velocity width may be due to galactic rotation in the extended
galaxies: the most spatially extended galaxies exhibit the broadest lines.    
\Ha\ traces the instantaneous star formation rate as the \Ha\ luminosity of a galaxy is directly proportional to the
ionising flux from massive stars, which drops off after only about 20
million years after star formation ceases. The UV luminosity, on the other hand, evolves
in time with changes in stellar population, continuing to rise
even after star formation has ended. We compared SFRs derived using the two
different methods, for each of our $>5\sigma$ detections, using conversions
given in Kennicutt (1998).
We performed photometry on the HST/ACS images from GOODS v1.0 (Giavalisco et al., 2003), with 1\arcsec\ apertures, for
consistency with our infra-red fibre size and used the B band magnitudes (4500\AA) to calculate rest frame UV (2400\AA) flux
densities and corresponding star formation rates. We find that SFRs calculated from the UV luminosity densities
are typically a factor of at least two lower than those derived from
\Ha. This is probably due to the differential effect of dust extinction
in the redshift one galaxies between $\lambda_{\rm rest}\approx$ 2400\AA\ and 6563\AA.
This result is consistent with values in Glazebrook et al. (1999),
Tresse et al. (2002), and Yan et al. (1999), who all
find SFR(\Ha)/SFR(UV) ratios of around 2-3.

\section{Summary}

We have performed multi-object, near-infrared spectroscopy of $z\sim1$
galaxies in the Hubble Deep Field North, using CIRPASS-MOS. These
observations form an important pilot study for an on-going survey to trace star formation
rates at redshift one. The brightest \Ha\ galaxies (about 10\% of our
sample) were detected at at least $5\sigma$ in only 3 hours exposure time, demonstrating the
success of this technique. We have shown that we can detect \Ha\ at
sufficient signal to noise to obtain a good estimate of star formation
rates in these galaxies.  By pre-selecting galaxies with H$\alpha$
redshifted between the OH sky lines, we can detect star formation
rates of $5\,h^{-2}_{70}\,M_{\odot}\,{\rm yr}^{-1}$ ($5\,\sigma$ in
3-hours). The star formation rates obtained are higher than those
estimated from UV continuum by a factor of about two, due to dust
obscuration in the UV. This is consistent with previous work in this
area. The results of this pilot study are discussed further in Doherty et
al (2004, submitted to MNRAS). The full survey will be presented in Doherty
et al. (2004b, {\it in prep}.) We have another $\sim60$ sources in this field, at known
redshifts, therefore by stacking the spectra we will be able to obtain
a global star formation rate for $z\sim1$ galaxies in the vicinity of
the HDF-N.  We aim to build up a sample
of several hundred galaxies in total, over several different fields, which will allow us to determine the \Ha\ 
luminosity function at $z\sim1$ and hence address the true star
formation rate at this important epoch.

\section*{Acknowledgments}
CIRPASS was built by the instrumentation group of the Institute of Astronomy in Cambridge, UK.
We thank the Raymond and Beverly Sackler Foundation and PPARC
for funding this project. We thank the WHT staff, in particular Danny Lennon, Kevin Dee, Rene Ruten,
Juerg Rey and Carlos Martin, for their help in enabling CIRPASS
to be used as a visitor instrument.  MD is grateful for support from the Fellowship
Fund Branch of AFUW Qld Inc., the Isaac Newton Studentship, the
Cambridge Commonwealth Trust and the University of Sydney. ECM acknowledges
the C.K. Marr Educational Trust.

\end{document}